\documentclass[11pt,reqno]{amsart}
\usepackage[latin1]{inputenc}
\usepackage{amsmath,latexsym,amssymb,graphicx,dsfont,amsthm,amsfonts}
\usepackage{color}
\usepackage{umoline}

\usepackage{natbib}

\usepackage{float}
\usepackage{hyperref}

\renewcommand{\O}{\mathbf{O}}

\newcommand{\elo}[1]{\mathrm{Elo}_{#1}}

\begin{document}

%
%
%

\numberwithin{equation}{section}

\title[Nested ZIGP Regression for FIFA World Cup 2022]{Nested Zero Inflated Generalized Poisson Regression for FIFA World Cup 2022}
\author{Lorenz A. Gilch}



\address{Lorenz A. Gilch: Universit\"at Passau, Innstrasse 33, 94032 Passau, Germany}

\email{Lorenz.Gilch@uni-passau.de}
\urladdr{http://www.math.tugraz.at/$\sim$gilch/}
\date{\today}

\begin{abstract}
This article is devoted to the forecast of the FIFA World Cup 2022 via nested zero-inflated generalized Poisson regression. Our regression model 
incorporates the Elo points of the participating teams, the location of the matches and the of team-specific skills in attack and defense as  covariates. The proposed model allows predictions in terms of probabilities in order to quantify the chances for each team to reach a certain stage of the tournament. We use Monte Carlo simulations for estimating the outcome of each single match of the tournament, from which we are able to simulate the whole tournament itself. The model is fitted on all football games of the participating teams since 2016 weighted by date and importance. Validation with previous tournaments and comparison with other Poisson models are given.
\end{abstract}

\maketitle

\section{Introduction}

Football is a typical low-scoring game and single matches are frequently decided through single events during the match. While several factors like extraordinary individual performances of single players or the whole team,  individual errors, injuries,  refereeing errors or just lucky coincidences are hard to forecast, each team has its strengths and weaknesses (e.g., attack and defense skills) and most of the results reflect these different qualities of the teams. Following this  idea we derive probabilities 
for the exact result of a single match between two participating teams, which involves the following four ingredients for both teams:
\begin{itemize}
\item Elo ranking
\item Attack strength 
\item Defense strength
\item Location of the match
\end{itemize}
The  complexity of the tournament with billions of different outcomes makes  it very difficult to obtain accurate estimates for the probabilities of certain events. Therefore, we do \textit{not} aim on forecasting the exact outcome of the tournament, but we want to make the discrepancy between the participating teams \textit{quantifiable} and we want to measure the chances of each team to reach certain stages of the tournament or to become world champion. In particular, since the groups are already drawn and the tournament structure for each team (in particular, the possible ways to the final) is set, the idea is to measure  whether a team has a rather simple or hard way to the final. 
\par
Let us give some background on modelling  results of football matches. Several statistical models have been proposed in the literature for the prediction of the outcome of football matches. They can be divided into two classes. The first class, the result-based model, models directly the probability of a game outcome in terms of win/loss/draw, while the second class, the score-based model, targets on the prediction of the exact match score. In this article we use the second approach since the match score is a non-neglecting, very important factor in the group phase of the world championship and it also implies a model for the first class. When two teams  have the same amount of points after having finished the group stage of the FIFA World Cup 2022 then the goal difference (and not the result of the direct duel!) determines which team will be ranked better than the other one. This may very likely result in a stronger opponent in the round of last $16$ for the worse ranked team. In fact,  as we have seen in former World Cups or in other tournaments, in most cases the goal difference is a crucial criterion which team will become group winner, runner-up or will be eliminated in the group stage. This observation underlines the importance and necessity of estimating the exact score of each single match and not only the outcome (win/loss/draw).
\par
There are several models for modelling the exact scores of football matches and most of them involve a Poisson model.  The easiest model assumes independence of the goals scored by each team and that each score can be modeled by a Poisson regression model; see, e.g., \cite{Le:97}.  Bivariate Poisson models, where the scores (i.e., the number of goals scored by both teams in a match) are modelled via two dependent Poisson distributed random variables,  were proposed by \cite{Ma:82} and extended by  \cite{DiCo:97} and  \cite{KaNt:03}. A short overview on different Poisson models  and related models like generalized Poisson models or zero-inflated models are given in \cite{ZeKlJa:08} and  \cite{ChSt:11}. Different similar models based on Poisson regression of increasing complexity (including discussion, goodness of fit and comparing them in terms of scoring functions) were analysed and used in \cite{gilch-mueller:18} for the prediction of the FIFA World Cup 2018. \cite{gilch:afc19} used a nested Poisson regression model for the Africa Cup of Nations 2019, Africa's continental championship, for predicting the hard-to-predict Africa Cup. Beyond the mentioned articles above forecast of football matches (in particular, Poisson regression based forecasts) has been studied in large variety also in recent years, amongst others \cite{Zeileis:21}, \cite{Wheatcroft}, \cite{azhari:18}, \cite{saraiva:16}. Let me also remark the work of \cite{JoZh:09} for a detailed comparison of generalized Poisson distribution and negative Binomial distribution, another class of possible distributions for modellung football scores.
\par
Possible covariates for the above models may be divided into two main categories: those  containing ``prospective'' informations and those containing ``retrospective'' informations. The first category contains other forecasts, especially bookmakers' odds, see e.g.  \cite{LeZeHo:10a}, \cite{LeZeHo:12} and references therein. This approach relies on the fact that bookmakers have a strong economic incentive to rate the result correctly and that they can be seen as experts in the matter of the forecast of sport events. However,  their forecast models remain undisclosed and rely on  information that is not publicly available.  
The second category contains only historical match data and no other forecasts. Models based on the second category allow to explicitly model the influence of objective, team-specific characteristics which will serve 
as covariates (in particular, strength and weakness in attack and defense). Although there is a lot of data collected during matches (e.g., pass accuracy, team's mileage, etc.), only a very limited number of data is publicly available which can be used as  possible  covariates for forecast models.  \cite{GrScTu:15} performed a variable selection on various covariates and found that the three most significant retrospective covariates are the FIFA ranking followed by the number of Champions league and Euro league players of a team. However, at the time of this analysis the composition and the line ups of the teams have not been announced and hence the number of Champions/Euro league players as covariates are not available. This is the reason why our model will be based on a team ranking only.
\par
As a quantitative measure of the participating team strengths in this article,  we use the \textit{World Football Elo ranking}, which is publicly available under
\begin{equation}\label{elo-website}
\texttt{http://en.wikipedia.org/wiki/World\_Football\_Elo\_Ratings}.
\end{equation}
We do \textit{not} make use of the FIFA ranking (which is a simplified Elo ranking since July 2018), because the calculation of the FIFA ranking changed over time and the Elo ranking is  more widely used in football forecast models. See also \cite{GaRo:16} for a discussion on this topic and a justification of the Elo ranking. Our model will be based on this Elo ranking and matches of the participating teams since 2016, where we additionally take the location of matches into account.
\par
In this article we follow  the retrospective approach and we present a nested generalized Poisson regression model with zero-inflation for the prediction of the scores of single matches, where the model is solely based on the Elo ranking and matches of the participating teams since 2016, where we additionally take the location of matches into account
Since the  FIFA World Cup 2022 is a complex tournament, involving important effects such as, e.g., group draws (e.g., see  \cite{De:11}) and dependences of the different matches,   Monte-Carlo simulations  are used to forecast the whole course of the tournament. For a more detailed summary on statistical modeling of major international football events, see, e.g.,  \cite{GrScTu:15}  and references therein. 
As we will see later our proposed model shows a good fit, the obtained forecasts are conclusive and outperform classical Poisson models in many cases; furthermore, our model gives \textit{quantitative insights} in each team's individual chances to proceed to certain stages of the tournament. 
 \par
The paper is organized as follows: in Section \ref{sec:model} we present our nested zero-inflated generalized Poisson regression model. In Section \ref{sec:model} we validate the proposed model by comparing the simulation results with the simulation results of classical Poisson regression models in view of tournaments from the past. The simulation results for the FIFA World Cup 2022 will be presented in Section \ref{sec:forecast}. Finally, in Section \ref{sec:discussion} we discuss our model and in Section \ref{sec:conclusion} we give concluding remarks.

\section{The Nested ZIGP Model}
\label{sec:model}

\subsection{Preliminaries}
\label{subsec:goals}
The simulation of the whole FIFA World Cup 2022 is based on the simulation of the \textit{exact} result of each single match which is modeled as $G_{A}$:$G_{B}$, where $G_{A}$ ($G_{B}$, respectively) is the number of goals scored by team A (by team B, respectively).  Hence, simulating each single match allows us to simulate the course of the whole tournament. Even the most probable tournament outcome has a probability very close to zero  to be actually realized. Hence, deviations of the true tournament outcome from the model's most probable one are not only possible, but most likely. However, simulations of the tournament yield estimates of the probabilities for each team to reach certain stages of the tournament and allow to make the different team's chances \textit{quantifiable}. 
\par
We are interested to give quantitative insights into the following questions:
\begin{enumerate}
\item Which team has the best chances to become new world champion?
\item How big are the probabilities that a team will win its group or will be eliminated in the group stage?
\item How big is the probability that a team will reach a certain stage of the tournament?
\end{enumerate}

\subsection{Involved data}

The main idea is to predict the exact outcome of a single match based on a generalized Poisson regression model which incorporates the following covariates:
\begin{itemize}
\item Elo ranking of the teams
\item Attack and defense strengths of the teams
\item Location of the match (either one team plays at home or the match takes place on neutral ground)
\end{itemize}
At this point let me remark that the national team of Qatar will be the single team which plays at home. However, for estimating the different attack and defense skills of all participating teams, we will take into account all matches since $2016$; of course, for estimating the regression parameters from historic match data it will be crucial whether a match was played on neutral playground or not.
\par
We use an Elo rating system, see  \cite{Elo:78}, which includes modifications to take various football-specific variables (like home advantage, goal difference, etc.) into account. The Elo ranking is published by the website \texttt{eloratings.net}, from where also all historic match data was retrieved.
We give a quick introduction to the formula for the Elo ratings: let $\mathrm{Elo}_{\mathrm{before}}$ be the Elo points of a team before a match; then the Elo points $\mathrm{Elo}_{\mathrm{after}}$  after the match against an opponent with Elo points $\mathrm{Elo}_{\mathrm{Opp}}$ is calculated as follows:
$$
\mathrm{Elo}_{\mathrm{after}}= \mathrm{Elo}_{\mathrm{before}} + K\cdot G\cdot (W-W_e),
$$
where 
\begin{itemize}
\item $K$ is a weight index regarding the tournament of the match (World Cup matches have weight $60$, while continental tournaments have weight $50$, etc.)
\item $G$ is a number taking into account the goal difference:
$$
G=\begin{cases}
1, & \textrm{if the match is a draw or won by one goal,}\\
\frac{3}{2}, & \textrm{if the match is won by two goals,}\\
\frac{11+N}{8}, & \textrm{where $N$ is the goal difference otherwise.}
\end{cases}
$$
\item $W$ is the result of the match: $1$ for a win, $0.5$ for a draw, and $0$ for a defeat.
\item $W_e$ is  the expected outcome of the match calculated as follows:
$$
W_e=\frac{1}{10^{-\frac{D}{400}}+1},
$$
where $D=\mathrm{Elo}_{\mathrm{before}}-\mathrm{Elo}_{\mathrm{Opp}}$ is the difference of the Elo points of both teams.
\end{itemize}
The Elo ratings  on 30 October  2022  for the top $6$ participating nations in the FIFA World Cup 2022 (in this rating) were as follows:
\begin{center}
\begin{tabular}{|c|c|c|c|c|c|}
\hline
Brazil &  Argentina & Spain & Netherlands & Belgium & France   \cr
\hline
2169 & 	2141    & 	2045 & 2040 & 2025	& 2005 \cr
\hline
\end{tabular}
\end{center}

Our model will be fitted  using  all matches of the participating teams  between 1 January 2016 and 30 October 2022. In addition, the historic match data is weighted according to the following criteria:
\begin{itemize}
\item Importance of the match according to FIFA weights
\item Time depreciation 
\end{itemize}
In order to weigh the historic match data for the regression model we use the following date weight function for a match $m$:
$$
w_{\textrm{date}}(m)=\Bigl(\frac12\Bigr)^{\frac{D(m)}{H}},
$$
where $D(m)$ is the number of days ago when the match $m$ was played and $H$ is the half period in days, that is, a match played $H$ days ago has half the weight of a match played today. Here, we choose the half period as $H=365\cdot 3\textrm{ days}= 3\,\textrm{years}$; compare with  \cite{Ley:19}.
\par
For weighing the importance of a match $m$, we use the match importance ratio in the FIFA ranking which is given by
$$
w_{\textrm{importance}}(m)=\begin{cases}
4, & \textrm{if $m$ is a World Cup match},\\
3, & \textrm{if $m$ is a continental championship/Confederation Cup match},\\
2.5, & \textrm{if $m$ is a World Cup or continental qualifier/Nations League match},\\
1, & \textrm{otherwise}.\\
\end{cases}
$$ 
The overall importance of a single match from the past will be assigned as
$$
w(m)= w_{\textrm{date}}(m) \cdot w_{\textrm{importance}}(m).
$$

\subsection{Nested Zero-Inflated Generalized Poisson Regression}
\label{sec:ZIGP}

We present a \textit{nested} regression approach for estimating the probabilities of the exact result of single matches. For this purpose,
we model the number of goals scored by a team in a single match as a random variable which follows 
a zero-inflated generalised Poisson-distribution (ZIGP). Generalised Poisson distributions generalise the Poisson distribution by adding a dispersion parameter; additionally, a point measure at $0$ is added, since the event that no goal is scored by a team typically is a special event. We recall the definition that a discrete random variable $X$ follows a \textit{Zero-Inflated Generalized Poisson distribution} with Poisson parameter $\mu>0$, dispersion parameter $\varphi\geq 1$ and zero-inflation $\omega\in[0,1)$ if
$$
\mathbb{P}[X=k]=\begin{cases}
\omega + (1-\omega) \cdot e^{-\frac{\mu}{\varphi}}, & \textrm{if k=0,}\\
(1-\omega)\cdot \frac{\mu\cdot \bigl(\mu+(\varphi-1)\cdot k\bigr)^{k-1}}{k!}\varphi^{-k} e^{-\frac{1}{\varphi}\bigl(\mu+(\varphi-1)x\bigr)}, & \textrm{if $k\in\mathbb{N}$};
\end{cases}
$$
compare, e.g., with  \cite{Co:89} and  \cite{St:04}. If $\omega=0$ and $\varphi=1$, then we obtain just the classical Poisson distribution. The advantage of ZIGP is now that we have an additional dispersion parameter. We also note that
\begin{eqnarray*}
\mathbb{E}(X) &=& (1-\omega)\cdot \mu, \\
\mathrm{Var}(X) &=& (1-\omega)\cdot \mu \cdot (\varphi^2 + \omega \mu).
\end{eqnarray*}
The idea is now to model the number  of scored goals of a team by a ZIGP distribution, whose parameters depend on the opponent's Elo ranking and the location of the match. Moreover, the number of goals scored by the weaker team (according to the Elo ranking) does additionally depend on the number of scored goals of the stronger team.
\par
We now explain the regression method in  detail.
Consider a match between two teams $A$ and $B$ whose result we want to estimate in terms of probabilities.
In the following we will always assume that $A$ has \textit{higher} Elo score than $B$. This assumption can be justified, since usually the better team dominates the weaker team's tactics. Moreover the number of goals the stronger team scores has an impact on the number of goals of the weaker team. For example,  if team $A$ scores  $5$ goals it is more likely that $B$ scores also $1$ or $2$ goals, because the defense of team $A$ lacks in concentration  due to the expected victory. If the stronger team $A$ scores only $1$ goal, it is more likely that $B$ scores no or just one goal, since team $A$ focusses more on the defense  and tries to secure the victory.
\par
Denote by $G_A$ and $G_B$ the number of goals scored by teams $A$ and $B$. Both $G_A$ and $G_B$ are assumed to be ZIGP-distributed: $G_A$ follows a ZIGP-distribution with parameters $\mu_{A|B}$, $\varphi_{A|B}$ and $\omega_{A|B}$, while $G_B$ follows a ZIGP-distribution with  parameters $\bar\mu_{B|A}$, $\bar\varphi_{B|A}$ and $\bar\omega_{B|A}$. These parameters are now determined as follows: 
\begin{enumerate}
\item The number of $G_A$ can be seen as the number $\hat G_{A}$ of scored goals from $A$ (strength of attack of $A$) or as the number $\check G_A$ of goals against of team $B$ (strength of defense of $B$). In the following we model $G_A$ from both points of view and take the average values of the parameters for modelling $G_A$.
\begin{enumerate}
\item In the first step we model the strength of team $A$ in terms of  the number of scored goals $\hat G_{A}$   in dependence of the opponent's Elo score $\elo{B}$ and the location of the match $\mathrm{loc}_{A|B}$, which is defined as
$$
\mathrm{loc}_{A|B}=\begin{cases}
1, & \textrm{if $A$ plays at home},\\
0, & \textrm{if the match takes place on neutral playground},\\
-1, & \textrm{if $B$ plays at home}.
\end{cases}
$$
The parameters of the distribution of $\hat G_A$ are modelled as follows:
\begin{equation}\label{equ:ZIGP-regression1}
\begin{array}{rcl}
\log \mu_A\bigl(\elo{B}\bigr)&=& \alpha_0^{(1)} + \alpha_1^{(1)} \cdot \elo{B} + \alpha_2^{(1)} \cdot \mathrm{loc}_{A|B},\\
\varphi_A &= &  1+e^{\beta^{(1)}}, \\
\omega_A &= & \frac{\gamma^{(1)}}{1+\gamma^{(1)}},
\end{array}
\end{equation}
where $\alpha_0^{(1)},\alpha_1^{(1)},\alpha_2^{(1)},\beta^{(1)},\gamma^{(1)}$  are obtained via ZIGP regression.  Recall again that $\hat G_A$ is a model for the scored goals of team $A$, which does \textit{not} take into account the specific defense skills of team $B$.

\item Teams of similar Elo scores  may have different strengths in attack and defense. To take this effect into account  we model the  number $\check G_A$ of goals team $B$ receives  against a team of higher Elo score  $\elo{A}$ using a ZIGP distribution with mean parameter $\nu_{B}$, dispersion parameter $\psi_B$ and zero-inflation parameter $\delta_B$ as follows:
\begin{equation}\label{equ:ZIGP-regression2}
\begin{array}{rcl}
\log \nu_B\bigl(\elo{A}\bigr) & =& \alpha_0^{(2)} + \alpha_1^{(2)} \cdot \elo{A} + \alpha_2^{(2)} \cdot \mathrm{loc}_{B|A},\\
\psi_B &= &  1+e^{\beta^{(2)}}, \\
\delta_B &= & \frac{\gamma^{(2)}}{1+\gamma^{(2)}},
\end{array}
\end{equation}
where $\alpha_0^{(2)},\alpha_1^{(2)},\alpha_2^{(2)},\beta^{(2)},\gamma^{(2)}$  are obtained via ZIGP regression. Recall that the number of scored goals of team $A$ can be seen as the goals against $\check G_A$ of team $B$.

\item Team $A$  shall in average score $(1-\omega)\cdot \mu_{A}(\elo{B})$ goals against team $B$ (modelled by $\hat G_A$), but team $B$ shall receive in average $(1-\omega_B)\cdot \nu_{B}(\elo{A})$ goals against (modelled by $\check G_A$). As these two values rarely coincide we model the numbers of goals $G_A$ as a ZIGP distribution with the average parameters
\begin{eqnarray*}
\mu_{A|B} &:= & \frac{\mu_A\bigl(\elo{B}\bigr)+\nu_B\bigl(\elo{A}\bigr)}{2},\\
\varphi_{A|B} &:=& \frac{\varphi_A + \psi_B}{2},\\
\omega_{A|B} &:= & \frac{\omega_A + \delta_B}{2}. 
\end{eqnarray*}
\end{enumerate}

\item The number of goals $G_B$ scored by $B$ is assumed to depend on the Elo score $\elo{A}$, the location $\mathrm{loc}_{B|A}$ of the match and additionally on the outcome of $G_A$. Hence, we model $G_B$ via a ZIGP distribution with Poisson parameters $\bar\mu_{B|A}$, dispersion $\bar\varphi_{B|A}$ and zero inflation $\bar\omega_{B|A}$ satisfying
\begin{equation}\label{equ:ZIGP-regression3}
\begin{array}{rcl}
\log \bar\mu_{B|A} &=& \alpha_0^{(3)} + \alpha_1^{(3)}  \cdot \elo{A}+\alpha_2^{(3)}\cdot \mathrm{loc}_{B|A} +\alpha_3^{(3)}  \cdot G_A,\\
\bar\varphi_{B|A} &:= &1+e^{\beta^{(3)}}, \\
\bar\omega_{B|A} &:= & \frac{\gamma^{(3)}}{1+\gamma^{(3)}},
\end{array}
\end{equation}
where the parameters $\alpha_0^{(3)},\alpha_1^{(3)} ,\alpha_2^{(3)} ,\alpha_3^{(3)} ,\beta^{(3)} ,\gamma^{(3)} $ are obtained by ZIGP regression. 

\item The result of the match $A$ vs. $B$ is simulated by realizing $G_A$ first and then  realizing $G_B$ in dependence of the realization of $G_A$. 
\end{enumerate}

\textbf{Example:} For better comprehension, we provide an example and consider the FIFA World Cup 2022 group match France vs. Denmark, which takes place in Qatar: France has $2005$ Elo points while Denmark has $1971$ points. Against a team of Elo score $1971$ France is assumed to score without zero-inflation in average
$$
\mu_{\textrm{France}}(1936)=\exp\bigl(2.472632      -0.0010679575 \cdot  1971 +0.2724600768 \cdot 0\bigr) = 1.444391
$$
goals, and France's zero inflation is estimated as
$$
\omega_{\textrm{France}} = \frac{e^{-4.058738}}{1+e^{-4.058738}}=0.0169776.
$$
Therefore, France is assumed to score in average
$$
(1-\omega_{\textrm{France}})\cdot \mu_{\textrm{France}}(1971) = 1.419869
$$
goals against Denmark. Furthermore, we obtain 
$$
\varphi_{\textrm{France}}= 1+ e^{-11.846181}=1.000007.
$$
Vice versa, Denmark receives in average without zero-inflation
$$
\nu_{\textrm{Denmark}}(2005)=\exp( -4.205890       +0.0021582919  \cdot 2005      -0.371076439 \cdot 0\bigr)= 1.129173
$$
goals, and the zero-inflation of Denmark's goals against is estimated as
$$
\delta_{\mathrm{Denmark}} = \frac{e^{-11.151799}}{1+e^{-11.151799}}=0.0000143.
$$
Hence,  Denmark receives in average
$$
(1-\delta_{\textrm{Denmark}})\cdot \nu_{\textrm{Denmark}}(2005) = 1.129157
$$
goals against when playing against an opponent of Elo strength $2005$. We obtain also
$$
\psi_{\textrm{Denmark}}= 1+ e^{-13.407282}=1.000002.
$$
Therefore,   the number of goals, which France will score against Denmark, will be modelled as a ZIGP distributed random variable with mean
$$
\mu_{\textrm{France}|\textrm{Denmark}}=\Bigl(1-\frac{\omega_{\textrm{France}}+\delta_{\mathrm{Denmark}}}{2}\Bigr)\cdot \frac{\mu_{\textrm{France}}(1971)+ \nu_{\textrm{Denmark}}(2005)}{2}=1.27585,
$$
dispersion parameter
$$
\varphi_{\textrm{France}|\textrm{Denmark}}= \frac{\varphi_{\textrm{France}}+\psi_{\textrm{Denmark}}}{2}=1.000005
$$
and zero-inflation
$$
\omega_{\textrm{France}|\textrm{Denmark}}= \frac{\omega_{\textrm{France}}+\delta_{\textrm{Denmark}}}{2}=0.008495965.
$$
The average number of goals, which Denmark scores against a team of Elo score $2005$ provided that $G_A$ goals against are received, is also modelled by a ZIGP distributed random variable with parameters
$$
\bar\mu_{\textrm{Denmark}|\textrm{France}} = \exp\bigl( 3.118465      -0.0013932201  \cdot 2005      -0.03989474 \cdot G_A  +     0.051954905\cdot 0\bigr);
$$
e.g., if $G_A=1$ then $\bar\mu_{\textrm{Denmark}|\textrm{France}}=1.32998$. Furthermore, we obtain
\begin{eqnarray*}
\bar\varphi_{\textrm{Denmark}|\textrm{France}} &=& 1+e^{-7.425663}=1.000596, \\
\bar\omega_{\textrm{Denmark}|\textrm{France}} &=& \frac{e^{-2.942621}}{1+e^{-2.942621}}=0.05008642.
\end{eqnarray*}
\textbf{Remark:} We note that the presented dependent approach may also be justified through the definition of  conditional probabilities:
$$
\mathbb{P}[G_A=i,G_B=j] = \mathbb{P}[G_A=i]\cdot \mathbb{P}[G_B=j \mid G_A=i] \quad \forall i,j\in\mathbb{N}_0.
$$
For a comparision of this model in contrast to similar Poisson models, we refer to \cite{gilch:afc19} and \cite{gilch-mueller:18}. All calculations were performed with R (version 4.0.3) and the \texttt{ZIGP} package. In particular, the presented model generalizes the model used in \cite{gilch:afc19} by adding a dispersion parameter, zero-inflation and a regression approach which weights historical data according to importance and time depreciation.

\subsection{Goodness of Fit Tests}\label{subsubsection:gof}
We check goodness of fit of  the ZIGP regressions in (\ref{equ:ZIGP-regression1}), (\ref{equ:ZIGP-regression2}) and (\ref{equ:ZIGP-regression3})  for all participating teams. For each team $\mathbf{T}$, we calculate the following  $\chi^{2}$-statistic from the list of historic matches:
$$
\chi_\mathbf{T} = \sum_{i=1}^{n_\mathbf{T}} \frac{(x_i-\hat\mu_i)^2}{\hat\mu_i},
$$
where $n_\mathbf{T}$ is the number of matches of team $\mathbf{T}$ since $2016$, $x_i$ is the number of scored goals of team $\mathbf{T}$ in match $i$ and $\hat\mu_i$ is the estimated ZIGP regression mean in dependence of the opponent's historical Elo points. 

\subsubsection{ZIGP Regression for $\hat G_A$}
\begin{figure}[H]
\centering
\includegraphics[width=4.8cm]{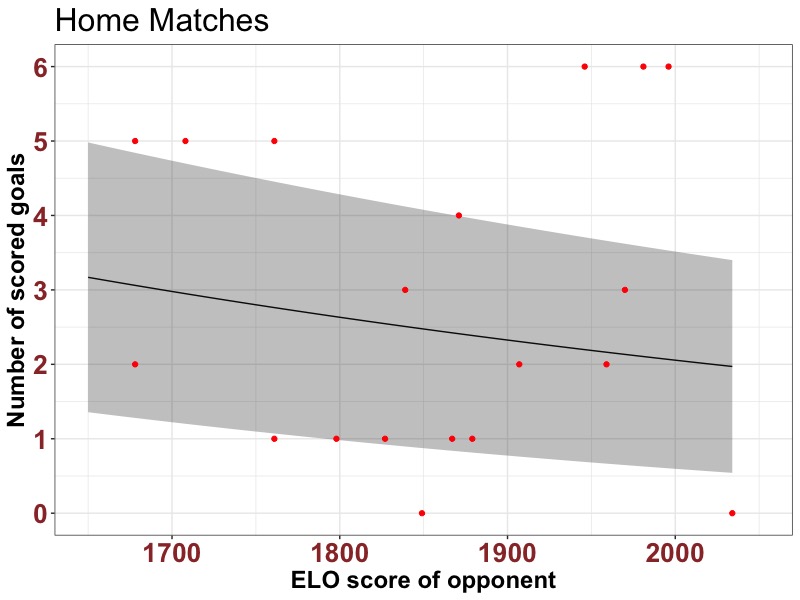}
\includegraphics[width=4.8cm]{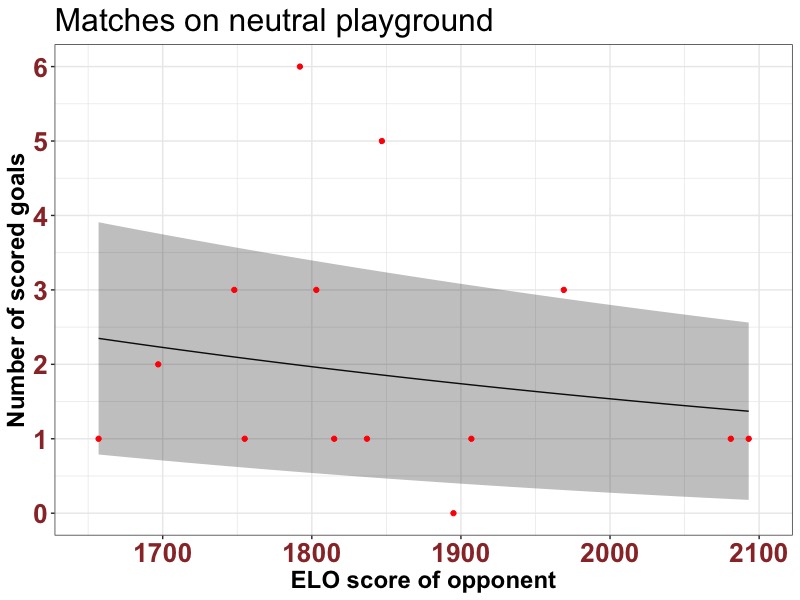}
\includegraphics[width=4.8cm]{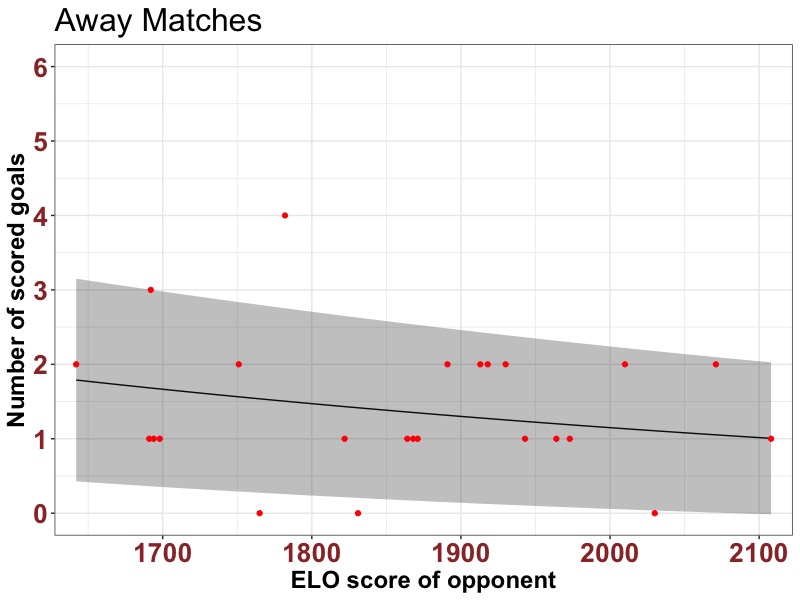}
\caption{ZIGP regression (\ref{equ:ZIGP-regression1}) for Spain.}
\label{fig:spain-regression1}
\end{figure}

Concerning the regression in (\ref{equ:ZIGP-regression1}) we observe that almost all teams have a very good fit. In Table \ref{table:godness-of-fit:gc} the $p$-values for  the top $6$ teams are given. 
\begin{table}[H]
\centering
\begin{tabular}{|l|c|c|c|c|c|c|}
  \hline
Team &  Brazil &  Argentina & Spain & Netherlands & Belgium & France  \\
\hline
$p$-value & 0.74 & 0.75 & 0.15 & 0.88 & 0.95 & 0.31 \\
   \hline
\end{tabular}
\caption{Goodness of fit test for the  ZIGP regression   in  (\ref{equ:ZIGP-regression1}) for the top $5$  teams. }
\label{table:godness-of-fit:gc}
\end{table}
Only Ghana and Costa Rica have a low $p$-value of less than $0.05$; all other teams have a $p$-value of at least $0.14$, most have an even much higher $p$-value. Since Costa Rica and Ghana belong to the weaker teams, these low $p$-values cause only limited effects on simulation results. The scored goals (red points) of Spain from historic match data together with the estimated averages $\mu_{\textrm{Spain}}$ (black line) and the standard error (shaded grey area) are given in Figure \ref{fig:spain-regression1}.

\subsubsection{ZIGP Regression for $\check G_A$}
\begin{figure}[H]
\centering
\includegraphics[width=4.8cm]{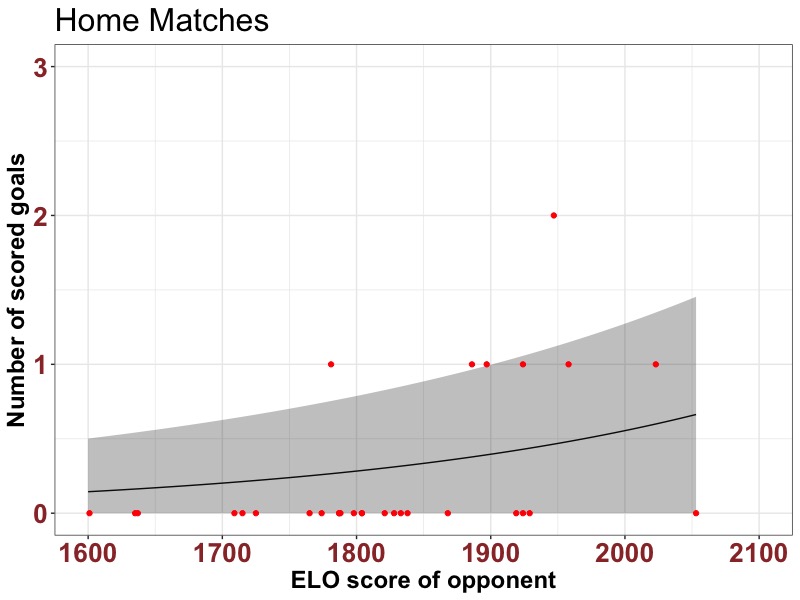}
\includegraphics[width=4.8cm]{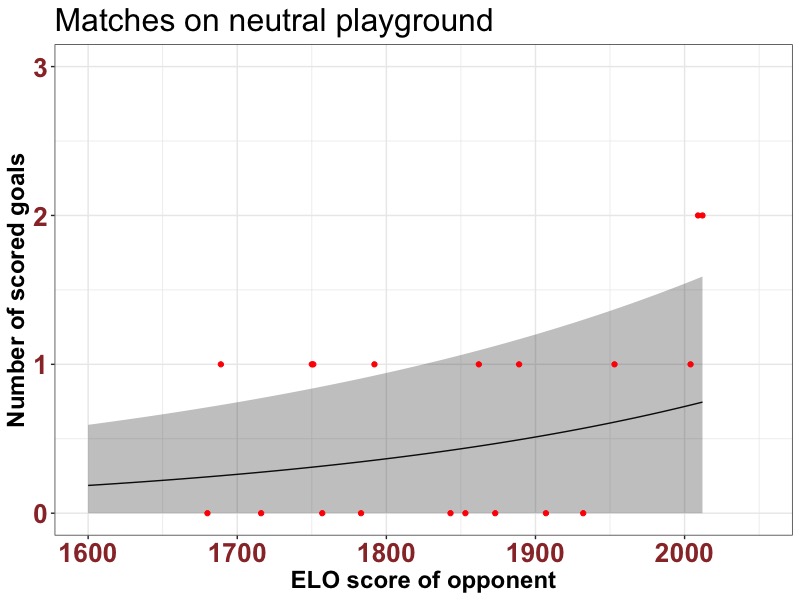}
\includegraphics[width=4.8cm]{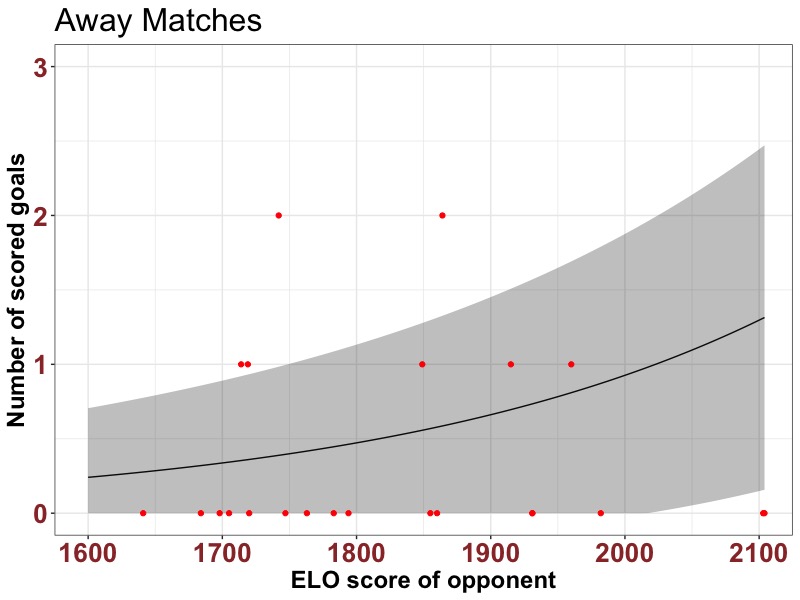}
\caption{ZIGP regression (\ref{equ:ZIGP-regression2}) for Brazil.}
\label{fig:brazil-regression2}
\end{figure}

Concerning the regression in  (\ref{equ:ZIGP-regression2}) which models the number of goals against we  calculate an analogous $\chi^{2}$-statistic for each team. The $p$-values for the top $6$ teams are given in Table  \ref{table:godness-of-fit2:gc}.
\begin{table}[H]
\centering
\begin{tabular}{|l|c|c|c|c|c|c|}
  \hline
  Team &  Brazil &  Argentina & Spain & Netherlands & Belgium & France  \\
  $p$-value & 0.69 & 0.005 & 0.77 & 0.82 & 0.16 & 0.86 \\
   \hline
\end{tabular}
\caption{Goodness of fit test for the  ZIGP regression   in  (\ref{equ:ZIGP-regression2}) for the top $5$ teams. }
\label{table:godness-of-fit2:gc}
\end{table}
Let us remark that three countries have a very poor $p$-value, namely Argentina, Canada and Uruguay ($0.032$); all other countries have $p$-values of at least $0.9$, most of them even much higher. Concerning Argentina's bad fit we remark that Argentina's regression (\ref{equ:ZIGP-regression1}) shows a good fit and limits the effect of Argentina's regression in (\ref{equ:ZIGP-regression2}). Furthermore, Canada  probably won't play a big role during the evolution of the tournament. The goals againt (red points) of Brazil from historic match data together with the estimated averages $\nu_{\textrm{Brazil}}$ (black line) and the standard error (shaded grey area) are given in Figure \ref{fig:brazil-regression2}.

\subsubsection{ZIGP Regression for $G_B$}

\begin{figure}[H]
\centering
\includegraphics[width=4.8cm]{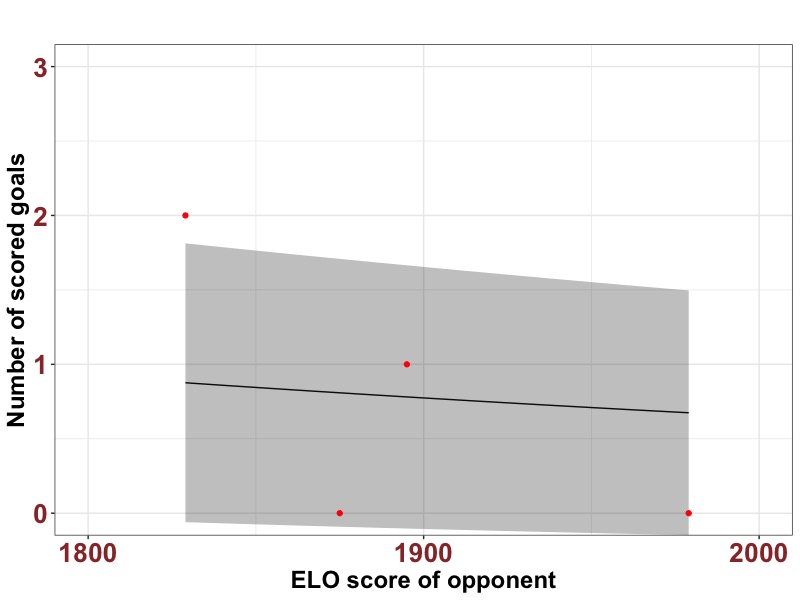}
\caption{ZIGP regression (\ref{equ:ZIGP-regression3}) for Wales' away matches against stronger opponents, who score exactly $2$ goals.}
\label{fig:wales-regression3}
\end{figure}

Finally, we test the goodness of fit for the regression in (\ref{equ:ZIGP-regression3}) which models the number of goals against of the weaker team in dependence of the number of goals which are scored by the stronger team; see Table \ref{table:godness-of-fit3:gc}. 
\begin{table}[H]
\centering
\begin{tabular}{|l|c|c|c|c|c|}
  \hline
 Team &  Spain & Argentina & Germany & Portugal  & Senegal 
 \\ 
  \hline
  $p$-value  &0.65 &  0.08 & 0.51 & 0.67  & 0.86 
    \\
   \hline
\end{tabular}
\caption{Goodness of fit test for the  Poisson regression   in  (\ref{equ:ZIGP-regression3}) for some of the  teams. }
\label{table:godness-of-fit3:gc}
\end{table}
Very few teams like Qatar or the United States have  $p$-values less than $0.05$, which arise mainly from very few outliers, while the $p$-values of the majority of the other teams suggest reasonable fits. For instance, the regression (\ref{equ:ZIGP-regression3}) for Wales, when playing away, facing a stronger opponent and when receiving exactly $2$ goals against, is plotted in Figure \ref{fig:wales-regression3}.

\subsection{Further Poisson Models}
\label{subsec:further-models}

In order to measure the quality of the proposed ZIGP model we compare it with other standard Poisson models. These are described in the following

\subsubsection{Independent Poisson Regression}

In this model we assume both $G_A$ and $G_B$ to be independent Poisson distributed variables with Poisson rates $\lambda_{A|B}$ and $\lambda_{B|A}$. We estimate the Poisson rates $\lambda_{A|B}$ and $\lambda_{B|A}$ via Poisson regression with  the Elo scores of $A$ and $B$ and the location $\mathrm{loc}_{A|B}$ of the match as covariates. These rates are determined as follows:
\begin{enumerate}
\item The first step models the number of goals $\hat G_{A}$ scored  by team $A$ playing against a team with a given Elo score $\elo{B}$. The random variable $\hat G_{A}$  is modeled as a Poisson distribution with parameter $\mu_{A}$. The parameter $\mu_{A}$ as a function of the Elo rating $\elo{B}$ of the opponent $B$ is given as
\begin{equation}\label{equ:independent-regression1}
\log \mu_A(\elo{B}) = \alpha_0 + \alpha_1 \cdot \elo{B} +\alpha_2 \cdot \mathrm{loc}_{A|B}\, 
\end{equation}
where $\alpha_0,\alpha_1$ and $\alpha_2$ are obtained via Poisson regression.  
\item Analogously to our ZIGP model, we model the number of scored goals of team $A$ also as the number of goals against of team $B$ when facing an opponent with Elo score  $\elo{A}$. We model this number as a Poisson distribution with parameter $\nu_{B}$ in terms of a function in the Elo rating $\elo{A}$ and the location of the match $\mathrm{loc}_{B|A}$:
\begin{equation}\label{equ:independent-regression2}
\log \nu_B(\elo{A}) = \beta_0 + \beta_1 \cdot \elo{A} + \beta_2 \cdot \mathrm{loc}_{B|A},
\end{equation}
where the parameters $\beta_0, \beta_1$ and $\beta_2$ are obtained via Poisson regression.
\item Team $A$  shall in average score $\mu_A\bigr(\elo{B}\bigr)$ goals against team $B$, but team $B$ shall receive in average $\nu_B\bigl(\elo{A}\bigr)$ goals against. Hence, we model the numbers of goals $G_A$ as a Poisson distribution with parameter
$$
 \lambda_{A|B} = \frac{\mu_A\bigl(\elo{B}\bigr)+\nu_B\bigl(\elo{A}\bigr)}{2}.
$$
Analogously, we set
$$
\lambda_{B|A} = \frac{\mu_B\bigl(\elo{A}\bigr)+\nu_A\bigl(\elo{B}\bigr)}{2}.
$$
\end{enumerate}
For each team, the regression parameters $\alpha_0, \alpha_1, \alpha_2, \beta_0,\beta_1$ and $\beta_2$ are estimated from historical match data. The match $A$ vs. $B$ is then  simulated using two \textit{independent} Poisson random variables $G_A$ and $G_B$ with rates $\lambda_{A|B}$ and $\lambda_{B|A}$.

\subsubsection{Bivariate Poisson Regression}

A weakness of the independent Poisson model is that the number of goals $G_A$ and $G_B$ are realized independently, which is a rather less reasonable assumption.
Therefore, we consider also a bivariate Poisson regression approach. First, recall the definition of a \textit{bivariate Poisson distribution}: let $X_1,X_2,X_0$ be \textit{independent} Poisson distributed random variables with rates $\lambda_1,\lambda_2,\lambda_0$. Define $Y_1=X_1+X_0$ and $Y_2=X_2+X_0$. Then $(Y_1,Y_2)$ is \textit{bivariate} Poisson distributed with parameters $(\lambda_1,\lambda_2,\lambda_0)$. In particular, $Y_i$ is Poisson distributed with rate $\lambda_i+\lambda_0$ and $\mathrm{Cov}(Y_1,Y_2)=\lambda_0$.
\par
In the following we model $(G_A,G_B)$ as a bivariate Poisson distributed random vector for every couple $A,B$ separately. The main idea is to perform one regression over all matches of team $A$ and to estimate the average number of scored goals of team $A$ and its opponent in terms of his Elo strength $\elo{B}$. Then we  perform another regression over all matches of team $B$ and  estimate the expected number of goals of $B$ and the average goals against of $B$ when playing against a team of Elo strength $\elo{A}$. Hereby, we use the same notation as in previous models with $\mu_{\mathbf{T}}$ being the Poisson rate for the average of scored goals of any team ${\mathbf{T}}$, while $\nu_{\mathbf{T}}$ is the average number of goals against of team ${\mathbf{T}}$.
The model  uses the following regression approach:
\begin{enumerate}
\item
For {each} World Cup participating team $\mathbf{T}$, we estimate the parameters $$(\lambda_1,\lambda_2, \lambda_0)=(\mu_{\mathbf{T}},\nu_{\mathbf{T}},\tau_{\mathbf{T}})$$  from the viewpoint of team $\mathbf{T}$, where we only take into account the matches of team $\mathbf{T}$. The parameters shall depend on the Elo strength $\elo{\O}$ of an opponent team $\O$ and on the location $\mathrm{loc}_{\mathbf{T}|\mathbf{O}}$ of the match. To this end, we use the following Poisson regression model:
\begin{eqnarray}
\log \mu_{\mathbf{T}}\bigl(\elo{\O}\bigr) &=& \alpha_{1,0} +\alpha_{1,1}\cdot \elo{\O}+\alpha_{1,2}\cdot \mathrm{loc}_{\mathbf{T}|\mathbf{O}},\nonumber\\
\log \nu_{\mathbf{T}}\bigl(\elo{\O}\bigr) &=& \alpha_{2,0} +\alpha_{2,1}\cdot \elo{\O}+\alpha_{2,2} \cdot \mathrm{loc}_{\mathbf{T}|\mathbf{O}},\label{equ:bivariate-regression1}\\
\log \tau_{\mathbf{T}}\bigl(\elo{\O}\bigr) &=& \alpha_{3,0} \nonumber
\end{eqnarray}
That is, the estimated expected number of scored goals of team $\mathbf{T}$ against a team of Elo strength $\elo{\O}$ at location $\mathrm{loc}_{\mathbf{T}|\mathbf{O}}$ is given by $\mu_{\mathbf{T}}\bigl(\elo{\O}\bigr)+\tau_{\mathbf{T}}$, while the estimated expected number of scored goals of a team with Elo score $\elo{\O}$ and location $\mathrm{loc}_{\mathbf{T}|\mathbf{O}}$ against $\mathbf{T}$ is given by $\nu_{\mathbf{T}}\bigl(\elo{\O}\bigr)+\tau_{\mathbf{T}}$. \par
\item In order to estimate the Poisson rates $(\lambda_1,\lambda_2,\lambda_0)$ for the match result $(G_A,G_B)$ we can use the regression coefficients both of $A$ and $B$ in the following way: 
 $\lambda_1$ may be estimated either by considering all matches of team $A$ and calculating $\mu_A\bigl(\elo{B}\bigr)$ or by considering all matches of team $B$ and calculating $\nu_B\bigl(\elo{A}\bigr)$, which corresponds to the goals against of team $B$ (that is, the number of scored goals of team $A$ against $B$). Therefore, we estimate $\lambda_1$ as the mean of $\mu_A\bigl(\elo{B}\bigr)$  and $\nu_B\bigl(\elo{A}\bigr)$. Analogously, we estimate $\lambda_2$ as the mean of $\mu_B\bigl(\elo{A}\bigr)$  and $\nu_A\bigl(\elo{B}\bigr)$ and $\lambda_0$ also as the mean of the covariances $\tau_A$ and $\tau_B$. That is,
\begin{eqnarray*}
\lambda_1 &=& \frac{\mu_A\bigl(\elo{B}\bigr)+ \nu_B\bigl(\elo{A}\bigr)}{2},\\
\lambda_2 &=& \frac{\mu_B\bigl(\elo{A}\bigr)+ \nu_A\bigl(\elo{B}\bigr)}{2},\\
\lambda_0 &=& \frac{ \tau_A\bigl(\elo{B}\bigr) +\tau_B\bigl(\elo{A}\bigr) }{2}.
\end{eqnarray*}
Finally, we assume that $(G_A,G_B)$ is bivariate Poisson distributed with parameters $(\lambda_1,\lambda_2,\lambda_0)$.
\end{enumerate}

\textbf{Remarks:} 
\begin{itemize}
\item In (\ref{equ:bivariate-regression1}) we estimate $\tau_{\mathbf{T}}$ to be a constant for each team. If we add $\elo{\O}$ as an additional covariate for $\tau_{\mathbf{T}}$, that is,
$$
\log \tau_{\mathbf{T}}\bigl(\elo{\O}\bigr) = \alpha_{3,0}+\alpha_{3,1}\elo{\O},
$$
then the AIC increases  for most of the teams, and thus we omit  $\elo{\O}$ as covariate.
\item For the simulation, we used the \texttt{bivpois}-package of Karlis and Ntzoufras.
\end{itemize}

\section{Validation of the ZIGP Model}
\label{subsec:validation}

In this section we want to compare the forecast quality of our nested ZIGP model with the standard Poisson models from Subsection \ref{subsec:further-models}. For this purpose, we made simulations for the past FIFA World Cups $2010$, $2014$ and $2018$  and the EURO $2016$ and $2020$, Europe's continental championship, and compared the obtained forecast with the real outcomes of the those tournaments. 

\subsection{Scoring Functions}
In order to compare simulations results with the real outcomes of tournaments in the past we use standard scoring functions . We introduce the following notation: let $\mathbf{T}$ be a participating national team. Then define:
$$
\mathrm{result}(\mathbf{T}) = \begin{cases}
1, &\textrm{if } \mathbf{T} \textrm{ became champion}, \\
2, &\textrm{if } \mathbf{T} \textrm{ went to the final but didn't win the final},\\
3, &\textrm{if } \mathbf{T} \textrm{ went to the semifinal but didn't win the semifinal},\\
4, &\textrm{if } \mathbf{T} \textrm{ went to the quarterfinal but didn't win the quarterfinal},\\
5, &\textrm{if } \mathbf{T} \textrm{ went to the round of last 16 but didn't win this round},\\
6, &\textrm{if } \mathbf{T} \textrm{ went out of the tournament after the round robin}.
\end{cases}
$$
E.g., for the World Cup $2014$ we obtain $\mathrm{result(Germany)}=1$, $\mathrm{result(Argentina)}=2$, or $\mathrm{result(Italy)}=6$. For every World Cup participant $\mathbf{T}$ we set the simulation result probability as $p_i(\mathbf{T}):=\mathbb{P}[\mathrm{result}(\mathbf{T})=i]$.

In order to compare the different simulation results with the reality we use the following well-known scoring functions:
\begin{enumerate}
\item \textbf{Brier Score:}
The error of team $\mathbf{T}$ is defined as
$$
\mathrm{error}_{\textrm{BS}}(\mathbf{T}) := \sum_{j=1}^6 \bigl( p_{j}(\mathbf{T}) -\mathds{1}_{[\mathrm{result}(\mathbf{T})=j]}  \bigr)^2.
$$
The total error score is then given by
$$
BS  = \sum_{\mathbf{T} \textrm{ tournament participant}} \mathrm{error}_{\textrm{BS}}(\mathbf{T}). 
$$ 
\item \textbf{Rank-Probability-Score (RPS):} 
The error of team $\mathbf{T}$ is defined as
$$
\mathrm{error}_{\textrm{RPS}}(\mathbf{T}) := \frac{1}{5} \sum_{i=1}^5 \left( \sum_{j=1}^i p_{j}(\mathbf{T}) -\mathds{1}_{[\mathrm{result}(\mathbf{T})=j]}  \right)^2.
$$
The total error score is then given by
$$
RPS  = \sum_{\mathbf{T} \textrm{ tournament participant}} \mathrm{error}_{\textrm{RPS}}(\mathbf{T}).
$$ 
\end{enumerate}

\subsection{Comparison of Different Models}

In  Table \ref{table:godness-of-fit2010:gc} we compare the quality of our nested ZIGP model forecast with the simulation results from standard Poisson regression models from Subsection \ref{subsec:further-models} in terms of the values of the scoring functions applied to  the World Cups $2010$, $2014$ and $2018$:

\begin{table}[H]
\centering
\begin{tabular}{|c|l|c|c|c|}
  \hline
 Year & Error function &  ZIGP   & BV & IP
 \\ 
  \hline
 2010 &  Brier Score & \textbf{17.79}  &17.97 & 17.97\\
  &Rank Probability Score &\textbf{4.93}  & 4.99& 5.05 \\
  \hline 
2014  &Brier Score & \textbf{20.18} & 22.33& 22.10 \\
  &Rank Probability Score & \textbf{5.06} & 5.52& 5.48\\
  \hline  
  2018 
 & Brier Score &18.51 & \textbf{18.00} & 18.39 \\
  &Rank Probability Score &5.54 &  \textbf{5.51} & \textbf{5.51}\\
   \hline
\end{tabular}
\caption{Validation of ZIGP models compared with other Poisson regression models measured by different scoring functions for the World Cups $2010$, $2014$ and $2018$ and the EURO $2016$ and $2020$. (BV=Bivariate Poisson, IP=Independent Poisson) }
\label{table:godness-of-fit2010:gc}
\end{table}

While our proposed ZIGP model clearly outperforms the standard Poisson regression model for the World Cups $2010$ and $2014$, the forecast for the World Cup $2018$ is worse than the standard Poisson models. A closer investigation shows that the bad, unpredictable performance of Germany (eliminated in the group stage with rather weaker teams after having finished the World Cup qualifiers with $10$ out of $10$ victories) had an heavy impact on the values of the scoring function. Furthermore, the ZIGP models still leads to similar  values (compared to the values of the standard Poisson models) of the scoring functions for the World Cup $2018$.

In Table \ref{table:godness-of-fit-EURO} we compare the quality of our ZIGP model forecast with the simulation results of standard Poisson models  for the EURO 2016 and 2020, Europe's continental championships. 

\begin{table}[H]
\centering
\begin{tabular}{|c|l|c|c|c|}
  \hline
 Year & Error function &  ZIGP   & BV & IP
 \\ 
  \hline
 2016 
  &Brier Score & \textbf{17.52}  &23.27 & 23.25\\
  &Rank Probability Score &\textbf{5.28}  & 5.97& 5.98 \\
  \hline 
2020   &Brier Score & 14.54 & \textbf{14.36} & 16.37 \\
  &Rank Probability Score & \textbf{5.06} & 5.07& 5.19\\
    \hline
\end{tabular}
\caption{Validation of ZIGP model compared with other Poisson regression models measured by different scoring functions for EURO $2016$ and $2020$. (BV=Bivariate Poisson, IP=Independent Poisson) }
\label{table:godness-of-fit-EURO}
\end{table}

For the EURP $2016$, we observe that our proposed ZIGP model clearly outperfoms the standard Poisson models. Concerning the EURO $2020$, the forecast of the bivariate Poisson model is very slightly better than our ZIGP model w.r.t. the Brier Score, but not w.r.t. the rank probability score; the difference in quality of the forecasts seems to be non significant.
\par
As a summary, saying that we do not expect to find a model which outperform other models in \textit{each and every} case  our proposed nested ZIGP model seems to be a reasonable improvement of standard Poisson models or may serve as an alternative model.
\par
Finally, let me remark that the proposed \textit{Nested Poisson Regression model} from \cite{gilch:afc19} and \cite{gilch-mueller:18} is in most cases also outperformed by the nested ZIGP model presented in this article.

\section{FIFA World Cup 2022 Forecast}
\label{sec:forecast}

In this section we present the simulation results for the FIFA World Cup 2022, which allows us to answer the questions formulated in Section \ref{subsec:goals}. For this purpose, we have simulated each single match of the FIFA World Cup 2022 via the presented nested ZIGP model from Subsection \ref{sec:ZIGP}, which in turn allows us to simulate the whole FIFA World Cup 2022 tournament. After each simulated match we have updated the Elo ranking according to the simulation results. This honours teams, which are in a good shape during a tournament and perform maybe better than expected. Overall, we have performed  $100.000$ simulations of the whole tournament with $\textsc{R}$, where we reset the Elo ranking at the beginning of each single tournament simulation. 

\subsection{Single Matches}

\begin{figure}[H]
\centering
\includegraphics[width=8cm]{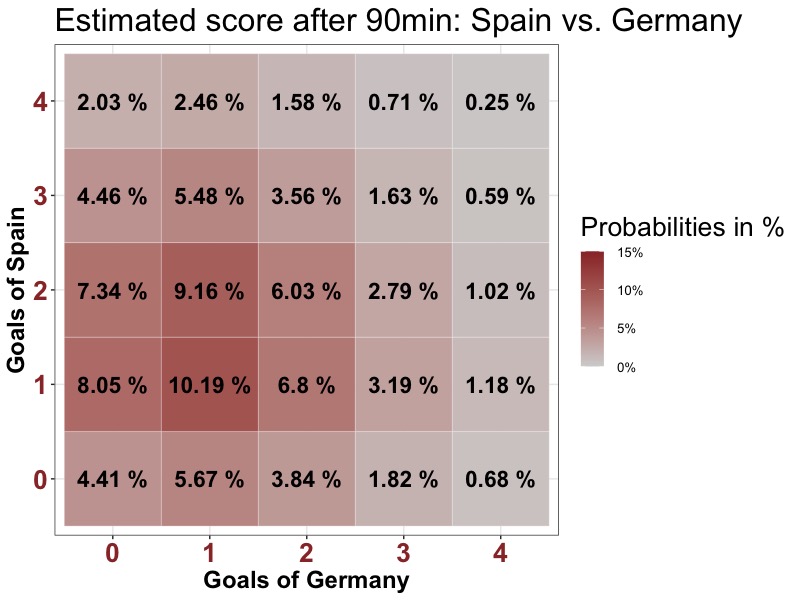}
\hspace{-0cm}\includegraphics[width=6.5cm]{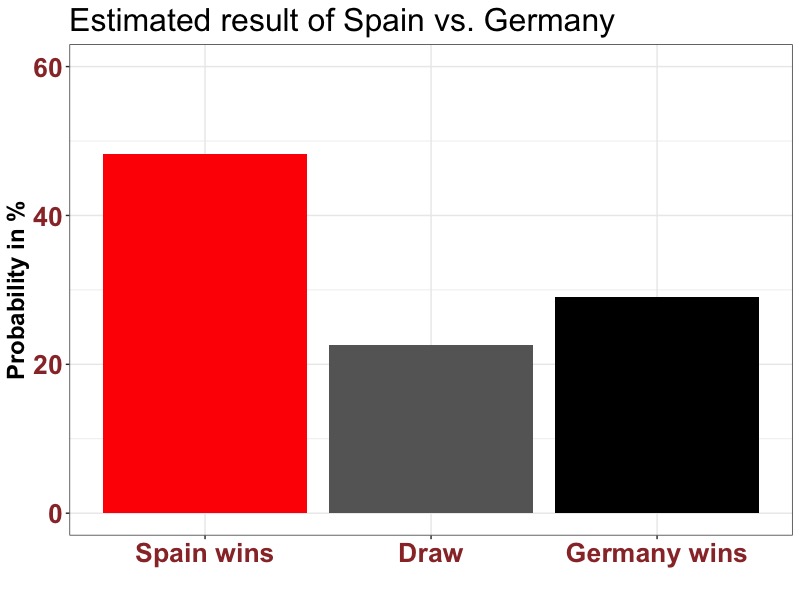}
\caption{Probabilities for the score of the preliminary round match Spain vs. Germany (group E) in Qatar.}
\label{table:ES-DE}
\end{figure}

Since the basic element of our tournament simulation is the simulation of single matches, we visualize how to quantify the results of single matches. E.g., in Group E the highly interesting match between Spain and Germany takes place in Qatar.  We present the probabilities  for the result of this match according  to our model in Figure \ref{table:ES-DE}: the most probable scores are a $1:1$ draw or a $2:1$ or $2:0$ victory of Spain. 

\subsection{Group Forecast} 

In the following Tables \ref{tab:WMgroupA}-\ref{tab:WMgroupH} we present the probabilities for the group stage that teams  become group winners, runner-ups, or are eliminated
 in the group stage. 

In most groups there is a clear favourite team but it is uncertain -- even among football experts -- which team will be runner-up. The forecast  quantifies and compares each team's chances to proceed to the round of last $16$ in terms of probabilities.

In Group E the top match of the group stage will take place between Spain and Germany. Although it is not unlikely that the direct match between both teams ends with a draw, the forecasts prefers Spain as group winner due to Spain's strength in attack, which makes it more likely for Spain to win the group also by goal difference.

\begin{table}[H]
\centering
\begin{tabular}{|r|ccc|}
  \hline
 Team & Group winner & Runner-up & Exit in preliminary round \\ 
  \hline
Netherlands & 61.2 \% & 23.2 \%& 15.6 \% \\ 
 Ecuador & 17.6 \%& 31.6 \% & 50.8 \% \\ 
 Senegal & 15.3 \%& 28.8 \% & 77.8 \%\\ 
 Qatar & 5.8 \%& 16.4 \%& 78.8 \% \\ 
   \hline
\end{tabular}
\caption{Probabilities for Group A}
\label{tab:WMgroupA}
\end{table}

\begin{table}[H]
\centering
\begin{tabular}{|r|ccc|}
  \hline
  Team & Group winner & Runner-up & Exit in preliminary round \\ 
  \hline
England & 54.7 \%& 24.5 \% & 20.9 \% \\ 
Wales & 18.8 \% & 29.1 \% & 52.1 \% \\ 
Iran & 13.8 \% & 23.8 \% & 62.3 \% \\ 
United States & 12.6 \% & 22.5 \%& 64.9 \% \\ 
   \hline
\end{tabular}
\caption{Probabilities for Group B}
\label{tab:WMgroupB}
\end{table}

\begin{table}[H]
\centering
\begin{tabular}{|r|ccc|}
  \hline
  Team & Group winner & Runner-up & Exit in preliminary round \\ 
  \hline
Argentina & 75.2 \%  & 17.6 \% & 7.2 \% \\ 
 Poland& 13.9 \% & 37.9 \% & 48.2 \% \\ 
  Mexico & 8.4 \% & 31.3 \% & 60.4 \% \\ 
  Saudi Arabia & 2.4 \% & 13.2 \% & 84.3 \% \\ 
   \hline
\end{tabular}
\caption{Probabilities for Group C}
\label{tab:WMgroupC}
\end{table}

\begin{table}[H]
\centering
\begin{tabular}{|r|ccc|}
  \hline
  Team & Group winner & Runner-up & Exit in preliminary round \\ 
  \hline
France& 47.4 \% & 34.3 \% & 18.2 \% \\ 
  Denmark & 42.8 \% & 36.5 \% & 20.8 \% \\
  Tunisia & 5.2 \% & 15.1 \% & 79.8 \% \\  
  Australia & 4.7 \% & 14.2 \% & 81.3 \% \\ 
   \hline
\end{tabular}
\caption{Probabilities for Group D}
\label{tab:WMgroupD}
\end{table}

\begin{table}[H]
\centering
\begin{tabular}{|r|ccc|}
  \hline
 Team & Group winner & Runner-up & Exit in preliminary round \\ 
  \hline
Spain & 54.2 \% & 29.3 \% & 16.76 \% \\ 
  Germany & 34.7 \% & 38.5 \% & 26.8 \% \\ 
  Japan & 7.4 \% & 19.0 \% & 73.7 \% \\ 
  Costa Rica & 3.9 \% & 13.1 \% & 82.9 \% \\ 
   \hline
\end{tabular}
\caption{Probabilities for Group E}
\label{tab:WMgroupE}
\end{table}

\begin{table}[H]
\centering
\begin{tabular}{|r|ccc|}
  \hline
  Team & Group winner & Runner-up & Exit in preliminary round \\ 
  \hline
Belgium & 57.0 \% & 24.8 \% & 18.1\%\\ 
  Croatia & 24.4 \% & 34.5 \% & 41.2 \% \\ 
  Morocco & 11.4 \% & 22.7 \% & 65.8 \% \\ 
  Canada & 7.2 \% & 17.9 \% & 74.9 \% \\ 
   \hline
\end{tabular}
\caption{Probabilities for Group F}
\label{tab:WMgroupF}
\end{table}

\begin{table}[H]
\centering
\begin{tabular}{|r|ccc|}
  \hline
 Team & Group winner & Runner-up & Exit in preliminary round \\ 
  \hline
Brazil & 64.2  \% & 24.4  \% & 11.5 \% \\ 
 Switzerland & 17.7  \% & 37.8  \% & 44.3  \% \\ 
  Serbia & 17.6  \% & 34.3 \% & 48.2  \% \\ 
  Cameroon & 0.4  \% & 3.6  \% & 96.0  \% \\ 
   \hline
\end{tabular}
\caption{Probabilities for Group G}
\label{tab:WMgroupG}
\end{table}

\begin{table}[H]
\centering
\begin{tabular}{|r|ccc|}
  \hline
  Team & Group winner & Runner-up & Exit in preliminary round \\ 
  \hline
Portugal & 59.8 \% & 27.7 \% & 12.4 \% \\ 
  Uruguay & 30.0 \% & 44.3 \% & 25.8 \% \\ 
  South Korea & 9.3 \% & 23.7 \% & 67.0 \% \\ 
  Ghana & 0.8 \% & 4.3 \% & 94.8 \% \\ 
   \hline
\end{tabular}
\caption{Probabilities for Group H}
\label{tab:WMgroupH}
\end{table}

\subsection{Playoff Round Forecast}

Our simulations yield the following probabilities for each team to win the tournament or to reach certain stages of the tournament. The result is  presented in Table  \ref{tab:FIFA2022}. The ZIGP regression model  favors Brazil and Argentina,  followed by   Belgium, Spain and the Netherlands. The current world champion France belongs also to the co-favourites, but the remaining teams have significantly less chances to win the FIFA World Cup 2022.

\begin{table}[ht]
\centering
\begin{tabular}{|r|cccccc|}
  \hline
  Team &  Champion & Final & Semifinal & Quarterfinal &  Last 16 & Prelim. Round \\ 
  \hline
  Brazil & 17.3 \% & 27.3 \% & 43.5 \% & 67.8 \% & 88.6 \% & 11.5 \% \\ 
   Argentina & 13.1 \% & 22.1 \% & 39.7 \% & 60.2 \% & 92.8 \% & 7.2 \% \\ 
   Belgium & 10.8 \% & 19.3 \% & 31.1 \% & 51.6 \% & 81.8 \% & 18.1 \% \\ 
   Spain & 10.2 \% & 17.7 \% & 30.1 \% & 52.4 \% & 83.5 \% & 16.6 \% \\ 
   Netherlands & 10.0 \% & 17.9 \% & 33.9 \% & 61.3 \% & 84.4 \% & 15.6 \% \\ 
   France & 8.4 \% & 16.8 \% & 32.1 \% & 51.7 \% & 81.8 \% & 18.2 \% \\ 
   Portugal & 6.5 \% & 13.8 \% & 24.5 \% & 46.0 \% & 87.5 \% & 12.4 \% \\ 
   Denmark & 6.1 \% & 13.5 \% & 27.9 \% & 46.9 \% & 79.3 \% & 20.8 \% \\ 
    Germany & 3.6 \% & 8.8 \% & 17.8 \% & 39.1 \% & 73.2 \% & 26.8 \% \\ 
    England & 3.5 \% & 9.4 \% & 22.0 \% & 45.3 \% & 79.2 \% & 20.9 \% \\ 
    Switzerland & 2.1 \% & 5.8 \% & 12.5 \% & 28.6 \% & 55.6 \% & 44.3 \% \\ 
    Serbia & 2.0 \% & 5.1 \% & 11.1 \% & 25.9 \% & 51.8 \% & 48.2 \% \\ 
    Croatia & 1.7 \% & 4.7 \% & 10.5 \% & 25.9 \% & 58.9 \% & 41.2 \% \\ 
    Senegal & 1.0 \% & 2.8 \% & 8.1 \% & 21.0 \% & 44.1 \% & 55.9 \% \\ 
  Uruguay & 0.9 \% & 3.3 \% & 8.7 \% & 24.3 \% & 74.3 \% & 25.8 \% \\ 
   Ecuador & 0.9 \% & 2.9 \% & 9.0 \% & 24.1 \% & 49.3 \% & 50.8 \% \\ 
   Poland & 0.6 \% & 2.3 \% & 8.2 \% & 18.7 \% & 51.8 \% & 48.2 \% \\ 
   Morocco & 0.4 \% & 1.2 \% & 3.6 \% & 11.8 \% & 34.1 \% & 65.8 \% \\ 
   Wales & 0.2 \% & 1.2 \% & 5.0 \% & 18.1 \% & 47.9 \% & 52.1 \% \\ 
   Japan & 0.2 \% & 0.8 \% & 2.3 \% & 8.1 \% & 26.4 \% & 73.7 \% \\ 
   Iran & 0.1 \% & 0.6 \% & 3.1 \% & 12.8 \% & 37.6 \% & 62.3 \% \\ 
   United States & 0.1 \% & 0.4 \% & 2.6 \% & 11.5 \% & 35.2 \% & 64.9 \% \\ 
   Mexico & 0.1 \% & 0.8 \% & 4.1 \% & 10.9 \% & 39.7 \% & 60.4 \% \\ 
   Canada & 0.1 \% & 0.5 \% & 1.8 \% & 7.0 \% & 25.1 \% & 74.9 \% \\ 
   South Korea & 0.1 \% & 0.5 \% & 1.6 \% & 6.5 \% & 33.1 \% & 67.0 \% \\ 
   Costa Rica & 0.1 \% & 0.2 \% & 0.9 \% & 4.3 \% & 17.1 \% & 82.9 \% \\ 
   Australia & 0.0 \% & 0.2 \% & 1.3 \% & 4.3 \% & 18.9 \% & 81.3 \% \\ 
   Tunisia & 0.0 \% & 0.2 \% & 1.2 \% & 4.2 \% & 20.3 \% & 79.8 \% \\ 
   Qatar & 0.0 \% & 0.1 \% & 1.0 \% & 5.9 \% & 22.2 \% & 77.8 \% \\ 
   Saudi Arabia & 0.0 \% & 0.1 \% & 0.8 \% & 3.3 \% & 15.6 \% & 84.3 \% \\ 
   Cameroon & 0.0 \% & 0.0 \% & 0.1 \% & 0.6 \% & 4.0 \% & 96.0 \% \\ 
   Ghana & 0.0 \% & 0.0 \% & 0.0 \% & 0.3 \% & 5.1 \% & 94.8 \% \\ 
   \hline
\end{tabular}
\caption{FIFA World Cup 2022 simulation results for the teams' probabilities to proceed to a certain stage.}
\label{tab:FIFA2022}
\end{table}

\begin{table}[H]
\centering

\end{table}


%
%

\section{Discussion}
\label{sec:discussion}

In this section we want to discuss  the presented ZIGP model and related models. In literature, one can find a great variety of scientific research papers which model the outcome of football matches by Poisson regression models. Most of these models use an independent or a bivariate Poisson regression approach. In the independent case there is only one single parameter to estimate which models the mean and variance simultaneously, while in the bivariate case one has an additional parameter which models only the correlation between the numbers of goals scored by both teams in a match. Multiplicative mixtures of distributions may lead to overdispersion. Therefore, it is desirable to use models having a variance function which is flexible enough to deal with overdispersion and underdispersion. This was the starting point to use an approach which allows additional flexibility of the variance.
One natural model for this is the generalised Poisson model, which was suggested by  \cite{Co:89} and where we have an additional parameter for modelling the variance in each regression. Additionally, we added a zero-inflation parameter in order to take into account the special value of $0$ (that is, no goals are scored by a team during a match), which leads to a ZIGP distribution. This was the starting point of the present article. For more details on (zero-inflated) generalised Poisson regression we refer to  \cite{St:04} and  \cite{Er:06}. Of course, the ZIGP model we used is not the only natural candidate for modeling football matches, but it can be seen as another reasonable class of Poisson models which are suitable for forecasting results of football matches. 
\par
We now want to discuss our proposed model. As we have shown in Subsection \ref{subsec:validation} the proposed ZIGP model with weighted historical data seems to improve standard models in most cases. However, as we have seen for the World Cup $2018$ outlier behaviour of some teams may be less efficient detected due to heavy weights of historic match data from the recent past. Hence, a possible weakness of the model could be the weights of historical matches where we used a half period of three years (suggested by \cite{Ley:19}). But we have observed that sometimes non-weighting of historical match data does not necessarily lead to worse qualities of the forecast. Outperformance of some teams during a championship is still hard to detect.

The goodness of fit for the teams seems to be reasonable but can be improved certainly by adding additional parameters (e.g., dependence not only on the opponents Elo strength but also on the opponents continental origin, etc.). However, this would go beyond the scope of this article. 
\par
Our proposed ZIGP model combines one team's attack strength and the opponent's defense skills in order to model the number of scored goals of one team. The weaker team's number of scored goals depends even on the number of scored goals of the stronger team. In combination with the use of ZIGP distributions this is a novelty in modelling football scores. Once again, this approach can be seen as a first stage in developing more accurate models based on these ideas. The aim of this article is to strengthen the use of ZIGP models for forecasting football scores, but further work has to be done in order to get improved models based on ZIGP regression models.

\par
For further discussion on adaptions and different models, we refer once again to the discussion section in \cite{gilch-mueller:18} and \cite{gilch:afc19}.

\section{Conclusion}
\label{sec:conclusion}

A team-specific zero-inflated generalized Poisson regression model for the number of goals in football matches  facing each other in international tournament matches has been used for quantifying the chances of the teams participating in the FIFA World Cup 2022. 
This model includes the Elo points of the teams as covariates and the location of the matches. The regression is based on historical match data  since 2016. The fitted model was used for Monte-Carlo simulations  of the FIFA World Cup $2022$. According to our simulation, the Brazil (followed by Argentina) turns out to be the top favorite for becoming new world champion. The current world champion France belongs only to the co-favourites. Moreover, for each team probabilities of reaching different stages of the tournament are calculated. 

\noindent
A major part of the statistical novelty of the presented work lies in the construction of the \textit{nested} ZIGP regression model. 
Validation of the proposed ZIGP model on tournaments between $2010$ and $2020$ showed a good fit and an outperformance of standard Poisson models in many cases.

\bibliographystyle{agsm}
\bibliography{bib}

\end{document}